\documentclass[twoside,twocolumn,9pt]{article}
\usepackage{extsizes}
\usepackage[super,sort&compress,comma]{natbib} 
\usepackage[version=3]{mhchem}
\usepackage[left=1.5cm, right=1.5cm, top=1.785cm, bottom=2.0cm]{geometry}
\usepackage{balance}
\usepackage{mathptmx}
\usepackage{sectsty}
\usepackage{graphicx} 
\usepackage{lastpage}
\usepackage[format=plain,justification=justified,singlelinecheck=false,font={stretch=1.125,small,sf},labelfont=bf,labelsep=space]{caption}
\usepackage{float}
\usepackage{fancyhdr}
\usepackage{fnpos}
\usepackage[english]{babel}
\addto{\captionsenglish}{%
  
}
\usepackage{array}
\usepackage{droidsans}
\usepackage{charter}
\usepackage[T1]{fontenc}
\usepackage[usenames,dvipsnames]{xcolor}
\usepackage{setspace}
\usepackage[compact]{titlesec}
\usepackage{hyperref}

\usepackage{graphicx}
\usepackage{amsmath}
\usepackage{bm}
\usepackage{setspace}
\usepackage{etoolbox}
\usepackage{makecell}
\usepackage{fancyhdr}
\usepackage{amssymb}
\usepackage{hyperref}
\usepackage{array}

\usepackage{epstopdf}

\definecolor{cream}{RGB}{222,217,201}
\newcommand{\minus}{\scalebox{0.75}[1.0]{$-$}}

\begin{document}

\pagestyle{fancy}
\thispagestyle{plain}
\fancypagestyle{plain}{
\renewcommand{\headrulewidth}{0pt}
}

\makeFNbottom
\makeatletter
\renewcommand\LARGE{\@setfontsize\LARGE{15pt}{17}}
\renewcommand\Large{\@setfontsize\Large{12pt}{14}}
\renewcommand\large{\@setfontsize\large{10pt}{12}}
\renewcommand\footnotesize{\@setfontsize\footnotesize{7pt}{10}}
\makeatother

\renewcommand{\thefootnote}{\fnsymbol{footnote}}
\renewcommand\footnoterule{\vspace*{1pt}%
\color{cream}\hrule width 3.5in height 0.4pt \color{black}\vspace*{5pt}} 
\setcounter{secnumdepth}{5}

\makeatletter 
\renewcommand\@biblabel[1]{#1}            
\renewcommand\@makefntext[1]%
{\noindent\makebox[0pt][r]{\@thefnmark\,}#1}
\makeatother 
\renewcommand{\figurename}{\small{Fig.}~}
\sectionfont{\sffamily\Large}
\subsectionfont{\normalsize}
\subsubsectionfont{\bf}
\setstretch{1.125} 
\setlength{\skip\footins}{0.8cm}
\setlength{\footnotesep}{0.25cm}
\setlength{\jot}{10pt}
\titlespacing*{\section}{0pt}{4pt}{4pt}
\titlespacing*{\subsection}{0pt}{15pt}{1pt}

\fancyfoot{}
\fancyfoot[LO,RE]{\vspace{-7.1pt}\includegraphics[height=9pt]{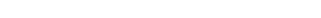}}
\fancyfoot[CO]{\vspace{-7.1pt}\hspace{13.2cm}\includegraphics{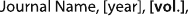}}
\fancyfoot[CE]{\vspace{-7.2pt}\hspace{-14.2cm}\includegraphics{head_foot/RF}}
\fancyfoot[RO]{\footnotesize{\sffamily{1--\pageref{LastPage} ~\textbar  \hspace{2pt}\thepage}}}
\fancyfoot[LE]{\footnotesize{\sffamily{\thepage~\textbar\hspace{3.45cm} 1--\pageref{LastPage}}}}
\fancyhead{}
\renewcommand{\headrulewidth}{0pt} 
\renewcommand{\footrulewidth}{0pt}
\setlength{\arrayrulewidth}{1pt}
\setlength{\columnsep}{6.5mm}
\setlength\bibsep{1pt}

\makeatletter 
\newlength{\figrulesep} 
\setlength{\figrulesep}{0.5\textfloatsep} 

\newcommand{\topfigrule}{\vspace*{-1pt}%
\noindent{\color{cream}\rule[-\figrulesep]{\columnwidth}{1.5pt}} }

\newcommand{\botfigrule}{\vspace*{-2pt}%
\noindent{\color{cream}\rule[\figrulesep]{\columnwidth}{1.5pt}} }

\newcommand{\dblfigrule}{\vspace*{-1pt}%
\noindent{\color{cream}\rule[-\figrulesep]{\textwidth}{1.5pt}} }

\makeatother

\twocolumn[
  \begin{@twocolumnfalse}
{\includegraphics[height=30pt]{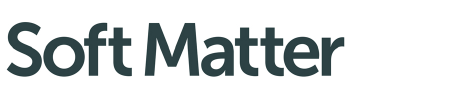}\hfill\raisebox{0pt}[0pt][0pt]{\includegraphics[height=55pt]{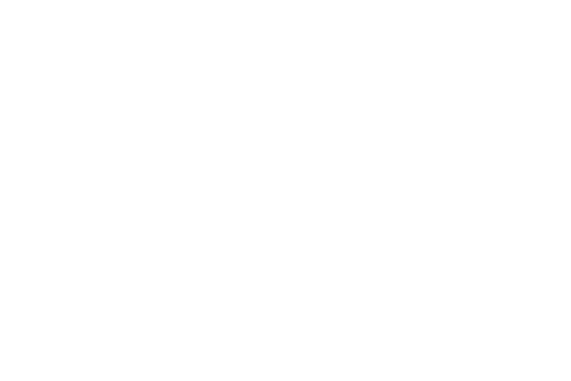}}\\[1ex]
\includegraphics[width=18.5cm]{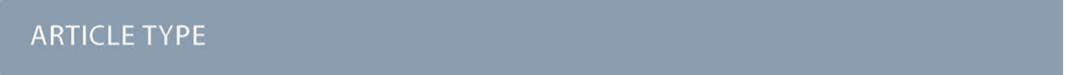}}\par
\vspace{1em}
\sffamily
\begin{tabular}{m{4.5cm} p{13.5cm} }

\includegraphics{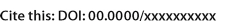} & \noindent\LARGE{\textbf{Faraday Cup Measurements of Triboelectrically Charged Granular Material: A Modular Interpretation Methodology$^\dag$}} \\
\vspace{0.3cm} & \vspace{0.3cm} \\

 & \noindent\large{Tom F. O'Hara$^{\ast}$, David P. Reid, Gregory L. Marsden, and Karen L. Aplin}\\

\includegraphics{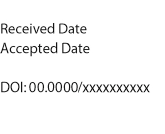} & \noindent\normalsize{The triboelectric charging of granular material is a long-standing and poorly understood phenomenon, with numerous scientific and industrial applications ranging from volcanic lightning to pharmaceutical production. The most widely utilised apparatus for the study of such charging is the Faraday cup, however existing analysis of the resulting measurements is largely simplistic and fails to distinguish charging due to particle-particle interactions from charging occurring through other mechanisms. In this contribution, we outline a modular approach for interpreting these measurements, enabling triboelectric phenomena to be explored in greater detail. In this approach, approximated charge distribution shapes are fitted to experimental Faraday cup traces. The fitting process uses measured size distributions in combination with simplified models of charge distribution and particle dynamics. This modular approach allows scope for adaptation of each step to fine-tune the process to specific application cases, making the technique broadly generalisable to any insulating granular material. An example case of volcanic ash showed that samples from the Gr\'imsv\"otn volcano charged with a greater proportion of particle-particle interactions than ash from Atitl\'an. Experimental validation is provided using sieved fractions of volcanic ash, where the broader size fractions were found to exhibit greater particle-particle charging. Non-particle-particle charging was also shown to scale with particle size as $\propto {d_p}^{\minus~0.85~\pm~0.03}$, roughly scaling with the particles' effective surface area.}\\

\end{tabular}

 \end{@twocolumnfalse} \vspace{0.6cm}

  ]

\renewcommand*\rmdefault{bch}\normalfont\upshape
\rmfamily
\section*{}
\vspace{-1cm}


\footnotetext{\textit{Faculty of Science and Engineering, University Walk, Bristol, BS8 1TR, UK. E-mail: tom.ohara@bristol.ac.uk}}

\footnotetext{\dag~Electronic Supplementary Information (ESI) available: [details of any supplementary information available should be included here]. See DOI: 10.1039/cXsm00000x/}


\section{Introduction}
Triboelectric phenomena are ubiquitous in both natural and industrial settings, influencing critical processes ranging from volcanic lightning and dust devils to planetary atmospheric electricity\cite{houghton_triboelectric_2013, reid_lab-based_2024, harrison_planetary_2008, rayborn_random_2022, cimarelli_volcanic_2022}. They also cause electrostatic ignition hazards in chemical processing and can reduce the powder-feeding performance of pharmaceuticals\cite{nfpa_recommended_2019, beretta_investigation_2020}. Yet despite the wide-ranging impacts of these phenomena, their underpinning mechanisms remain poorly understood. It is unclear whether the transfer of electrons, ions, or bulk material governs this process. Furthermore, efforts to unravel these phenomena are hindered by inconsistencies in how triboelectric charging of granular material is measured and interpreted\cite{lacks_long-standing_2019}.

The description of charge transfer is well understood for metals, compared to the insulators responsible for typically stronger triboelectric charging. On contact, metals transfer electrons according to their well-defined work functions. This results in a net negative charge remaining on the metal with a higher work function upon separation. However, the mechanisms behind triboelectric charging in insulators remain poorly understood. Proposed mechanisms include electron or ion transfer, mechanically induced bond breakages, and bulk material movement, each supported and refuted by varying evidence\cite{lacks_long-standing_2019, cezan_control_2019}. Some empirical understanding can be gained by placing materials in a ``Triboelectric Series'', in which the polarity of charge obtained from the rubbing of two materials can be predicted\cite{wilcke_dispvtatio_1757}. Unfortunately, this approach is highly system-dependent, with changes in environmental factors such as temperature and humidity changing the magnitude or even polarity of charging\cite{xu_experimental_2023, jallo_explaining_2015, biegaj_surface_2017, cruise_effect_2022, peart_powder_2001, cruise_triboelectric_2023}. It also fails to explain the charging between the same material, such as identical grains that have been shown to charge with a varying size dependence\cite{waitukaitis_size-dependent_2014}.

To better understand the system dependence of charge transfer in granular materials more reliable and well understood measurements are required. Faraday cups are the most widespread and commonly used apparatus for charge measurements of powders\cite{zhao_measurement_2002, houghton_triboelectric_2013, yeo_laboratory_2023, thomas_tribocharging_2008}. A Faraday cup is a typically cylindrical and shielded conductive vessel that exhibits ``electrification by induction'' as observed by Faraday in 1849\cite{maxwell_treatise_1873}. When a charged species impacts the cup, charge is transferred by conduction to the inner electrode, which can be measured using a connected electrometer\cite{kucerovsky_analysis_2003}. These charge measurements have a vast array of applications from ion beam characterisation and plasma diagnostics to the charged granular material relevant to this work\cite{sosolik_technique_2000, sunil_study_2023}. A basic Faraday cup can be employed to measure the net charge of a species landing in the cup, as in this work. Measurements can also be performed purely due to electrostatic induction in cases where particles do not make contact, commonly referred to as a ``Faraday pail''\cite{zhao_measurement_2002}. The charge carrier may also pass freely through the system, allowing for the measurement of a charged particle's velocity in a dynamic Faraday cup or be used as part of an array of cups to analyze subsets of the sample\cite{kucerovsky_analysis_2003}. To measure the transferred or induced charge electrometers can be used to detect the voltage or current and calculate the charge\cite{peart_powder_2001}.

Faraday cups are usually able to detect the polarity and magnitude of the net charge of powders, or subsets of sizes of the powder, but are not yet are unable to characterise the overall charge distribution with respect to size\cite{carter_effect_2020}. Other techniques are therefore under development to allow charge measurements to be resolved spatially or by size. One of the most promising alternative approaches is to use Particle Tracking Velocimetry (PTV), in which particles falling under gravity with an applied external electric field have their displacements tracked by a high-speed camera. Resolving Newtonian equations of motion yields the force experienced by the particles, allowing for the charge to be calculated\cite{waitukaitis_situ_2013}. Some of these techniques are not yet reliable, contain large uncertainties, and may have to be carried out under high vacuum\cite{carter_effect_2020}. However, these methods have shown that PTV can detect charges of granular materials up to 76 times greater than their average, which traditional Faraday cup methods would produce. This work aims to formulate a model in which contributions to Faraday cup charge measurements can be separated utilising the temporal resolution provided by powders falling at room temperature and pressure, using relatively low-cost apparatus where the conditions can be readily altered.

Existing techniques for analysis of powder traces from Faraday cups predominantly consist of taking the net difference before and after the charged granular material enters the cup, or by using the range of values acquired\cite{zhao_measurement_2002}. In some cases, the maximum may be assumed to be an asymptotic value when the smaller particles remain dispersed in the air as aerosol and may take far longer to deposit, if at all\cite{houghton_triboelectric_2013}. These measurements of net charge are quick and easy and can be carried out with individual particles or larger assemblies, however, the shape of the trace contains more information on the granular charging that could potentially be extracted than simply the maxima and minima\cite{yeo_laboratory_2023, houghton_triboelectric_2013}.

Taking a net charging approach yields a single value that, in many cases, is taken to indicate the relative extent of triboelectric charging, its polarity, or both. However, in cases where particle-particle charging is being investigated, this charge may also be due to contact with the container or from a preexisting charge distribution that could have been generated upon loading the sample into the respective measurement apparatus. Some works aim to study particle-particle charging in isolation by minimising particle-wall charging by reducing contact with container walls, which may be grounded, or by using a ``fountain-like'' flow in a low-pressure environment\cite{houghton_triboelectric_2013, forward_methodology_2009}. However, it is very difficult to entirely negate the effects of preexisting charges and particle-wall contacts. For instance, a granular sample with an initially positive trace from particle-particle charging and a negative charge from loading could yield a net result similar to or less than a case with less particle-particle charging but the same polarity as the non-particle-particle charge. A few example traces from samples of volcanic ash that will be used later in this work can be seen in Figure \ref{Example_Volcano_Traces} to demonstrate some of the different shaped traces that can be obtained. The goal of this paper is to develop a more sophisticated approach that feeds the entire Faraday cup trace into a model that predicts the extent of particle-particle charging separated from the other sources of charge, taking advantage of information stored in the overall shape of the trace.

In this work, the structure of a new modular approach for interpreting the Faraday cup measurements of triboelectrically charged granular material is outlined and applied. The experimental apparatus is detailed in Section \ref{Methodology} and then illustrated with example traces before specifying the numerical and simulation techniques employed at each step in the interpretation methodology. In Section \ref{Results and Discussion} the approach is initially applied to samples of volcanic ash, and then sieved fractions of ash are used to provide experimental validation for the new interpretation methodology, before investigating the scale-up of the non-particle-particle charging for simplified material. Finally, the work is concluded with a discussion of the limitations and potential application areas in Section \ref{Conclusions}.

\section{Methodology}
\label{Methodology}
\subsection{Model overview}
\label{Model Overview}

\begin{figure}[b!]
    \centering
    \includegraphics[height=6cm]{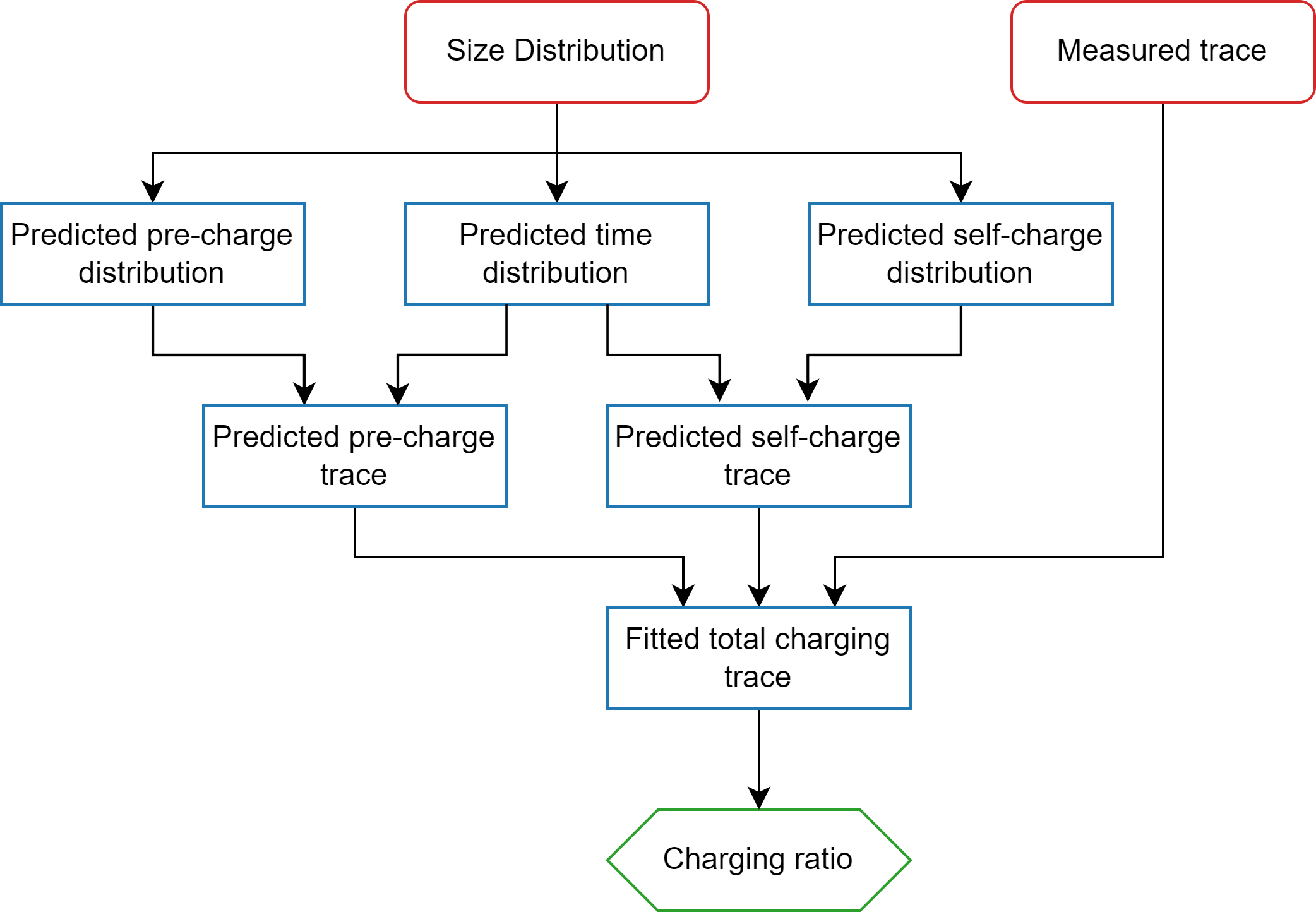}
    \caption{A visual schematic for the overall Faraday cup trace model. Red rounded rectangles represent the required inputs, the blue rectangles show the calculation and simulation steps and the green hexagon indicates the final output.}
    \label{Model_Overview}
\end{figure}

The approach taken in this work utilises the natural separation by size that occurs for granular materials (used here as a term encompassing powders and aerosols) falling under gravity in atmospheric conditions due to air resistance. For this modular approach, a size-dependent particle dynamics model for the particles' fall time was required, along with a size distribution of the particles and a prediction for the shape of the charge distribution function. These distributions were then combined to get the expected shape of the experimental trace. This process was done for both the particle-particle charging (often referred to in this work as self-charging) and the non-particle-particle charging arising from the loading of samples and charging with the container walls (pre-charging). The pre- and self-charging were then added to get the total charging. This total trace was then fitted to the experimentally acquired Faraday cup data by optimising for the relative ratios and polarities of the two types of charging. This procedure overall gives rise to predicted ratios of pre- and self-charging, such that the dependencies of the charging components on other factors can be investigated. An overall schematic for this model can be seen in Figure \ref{Model_Overview}.

In the examples used here, two charge distributions are defined, one for the pre-charging and one for the self-charging, each giving rise to an expected contribution to the final Faraday cup trace. However, the principle of this model is easily generalisable to a greater number of traces. For example, if the expected shape of the size dependencies for pre-charging arriving from different sources were known separately then the same model would, in principle, be able to resolve the relative charging arising from each source. 

\subsection{Experimental setup}

\begin{figure}[b!]
    \centering
    \includegraphics[height=9cm]{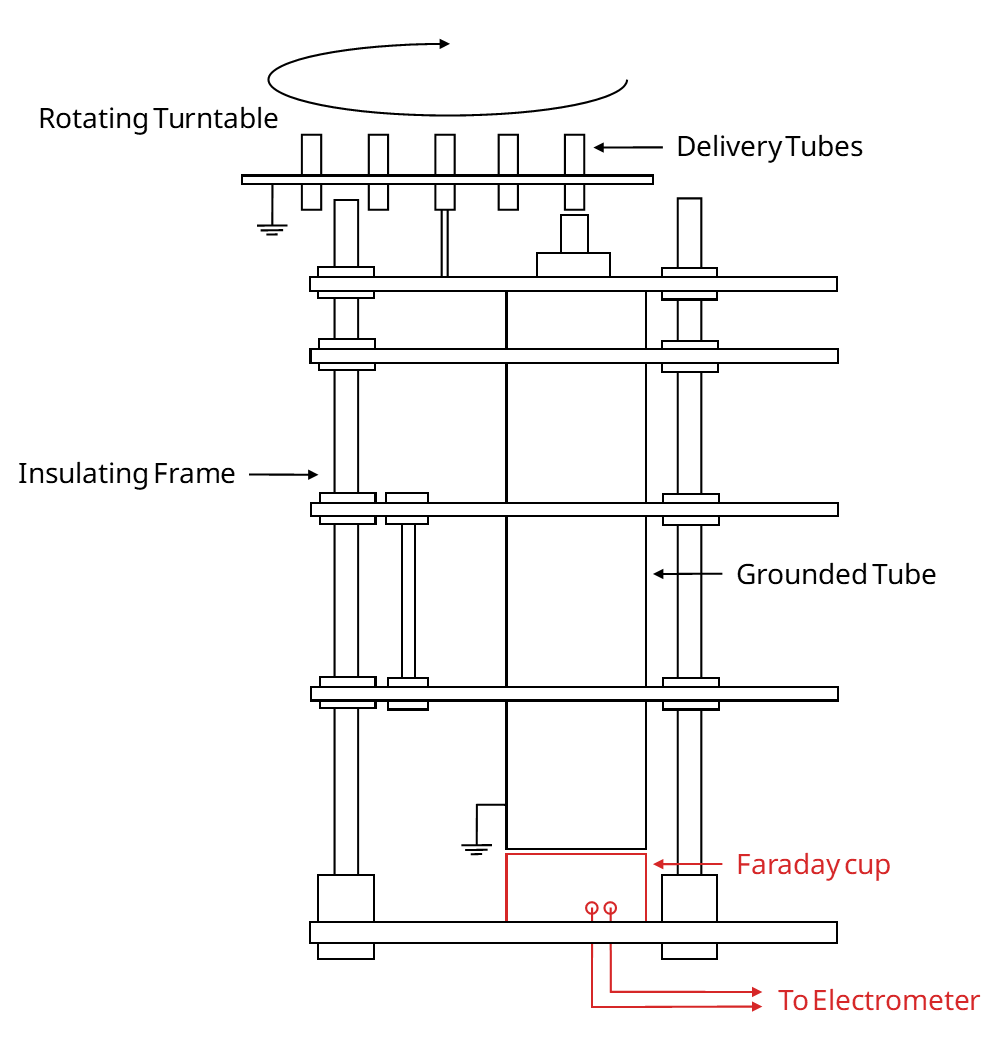}
    \caption{A schematic of the ash charge apparatus for this work, adapted from Houghton et al. with the Faraday cup highlighted in red\cite{houghton_triboelectric_2013}.}
    \label{Experimental_Setup}
\end{figure}

\begin{figure}[t!]
    \centering
    \includegraphics[height=5cm]{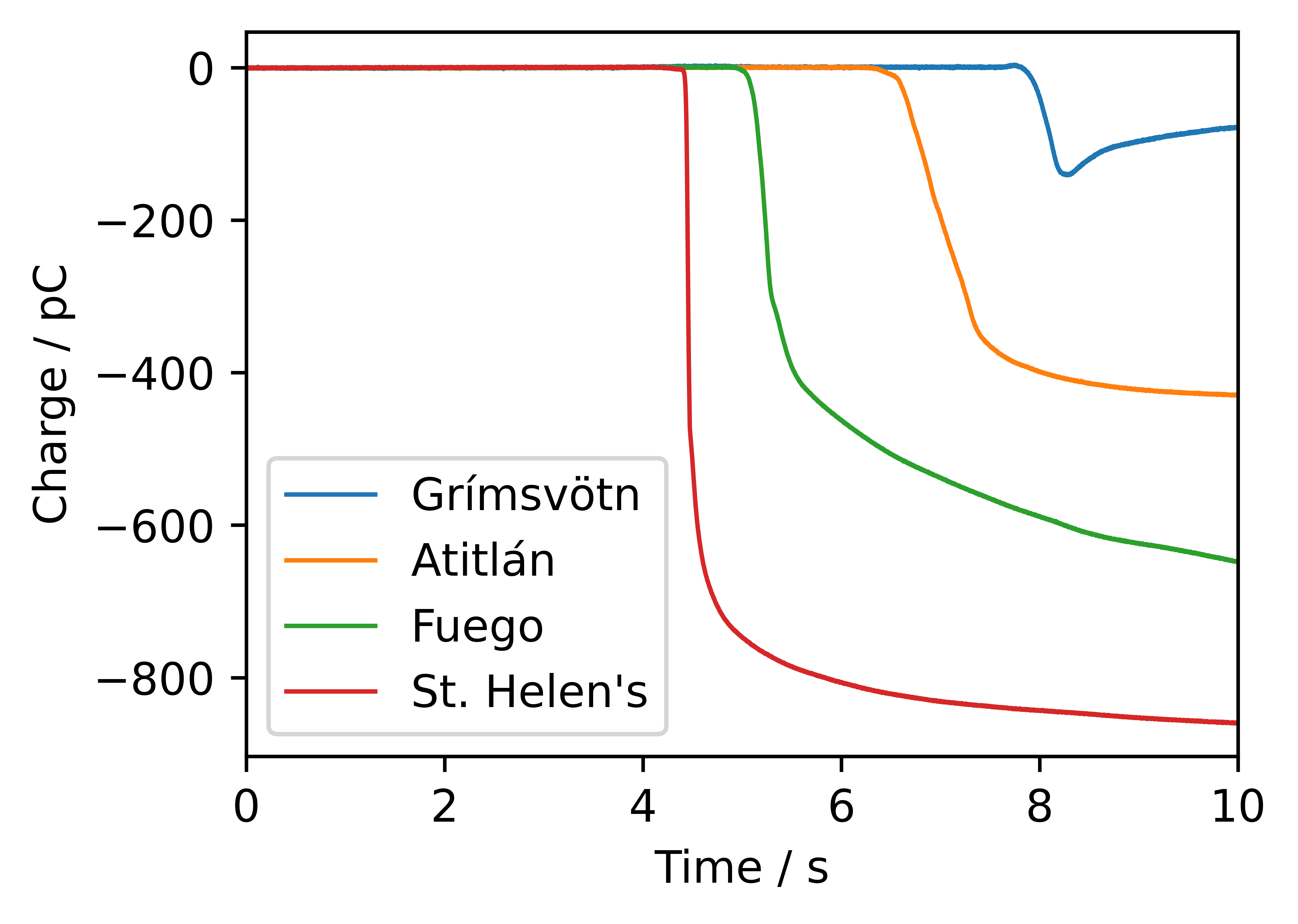}
    \caption{Example Faraday cup traces for samples produced from dropping samples of volcanic ash. Each trace is smoothed out by taking an average of multiple drops for different sub-samples, with each trace shown at staggered times for clarity.}
    \label{Example_Volcano_Traces}
\end{figure}

The experimental setup used in this work can be seen in Figure \ref{Experimental_Setup}. The basic principle is that granular samples can be loaded into the delivery tubes on top of the rotating turntable. The delivery tubes and turntable are grounded so that samples can be left for a period of time (usually around 20 minutes) in order for the majority of the residual charge, such as that obtained through loading, to decay\cite{houghton_triboelectric_2013}. The samples can then be dropped through a wider grounded tube, through two ring probes, before finally landing in the Faraday cup below. The charge transferred to the Faraday cup ($\Delta Q$) is then detected by the electrometer as a voltage ($\Delta V$). The measured voltage is related to the transferred charge and the capacitance ($C$) of the system by $\Delta V = \frac{\Delta Q}{C}$. The capacitance for this setup was previously determined by Houghton \textit{et al.} as 130~pF. This was done by measuring current ($I$) whilst applying a constant voltage change over time $\frac{dV}{dt}$, using the relation $I = C \frac{dV}{dt}$ to calculate the capacitance\cite{houghton_triboelectric_2013}. Overall, this methodology gives rise to traces of charge (or specific charge) over time, as the various charged particles enter the cup, as can be seen in Figure \ref{Example_Volcano_Traces}.

The falling section of this tube is fairly generalisable but the key quantity that will be used in Section \ref{Particle Dynamics Model} for the particle dynamics model is the drop height, which in this case was 37.25~cm. In the first implementation of this methodology any inductive effect before contact with the cup is assumed to be negligible. The cylindrical delivery tubes had heights of 13~mm and diameters of 7~mm. This means the available volume for a sample was 500~mm$^3$, hence the dropped samples had masses on the scale of grams. For each trace, the variation in mass can be adjusted for by dividing by the mass of the sample dropped, yielding the specific mass in each case if required. The Faraday cup was connected through BNC connectors to a \href{https://ilg.physics.ucsb.edu/Courses/RemoteLabs/docs/Keithley6514manual.pdf}{Keithley 6514 system electrometer}. Short rigid BNC connectors were utilised to minimise noise in the low-level measurements\cite{keithley_low_1998}. The output of the electrometer was plugged into a \href{https://www.ni.com/docs/en-US/bundle/usb-6211-specs/page/specs.html}{USB-6210} Data Acquisition (DAQ) device from National Instruments that was processed using \href{https://www.ni.com/en/support/downloads/software-products/download.labview.html}{LabVIEW$^{\circledR}$ 2024-Q1} before post-processing with the outlined model. The DAQ was also connected to a \href{https://datasheet.octopart.com/386-Adafruit-Industries-datasheet-81453130.pdf}{DHT11 humidity and temperature sensor} via an \href{https://docs.arduino.cc/resources/datasheets/A000067-datasheet.pdf}{Arduino Mega 2560} to record the environmental conditions of each drop. The temperature and humidity were not controlled but were recorded for reference at the time of each drop with common values around 24~°C and 40\%, respectively. Over a 2~hr measuring period typical variations of around  1~°C and 2\% relative humidity were observed.

\subsection{Size distribution fitting}
\label{Size Distribution Fitting}
\begin{figure}[b!]
    \centering
    \includegraphics[height=12cm]{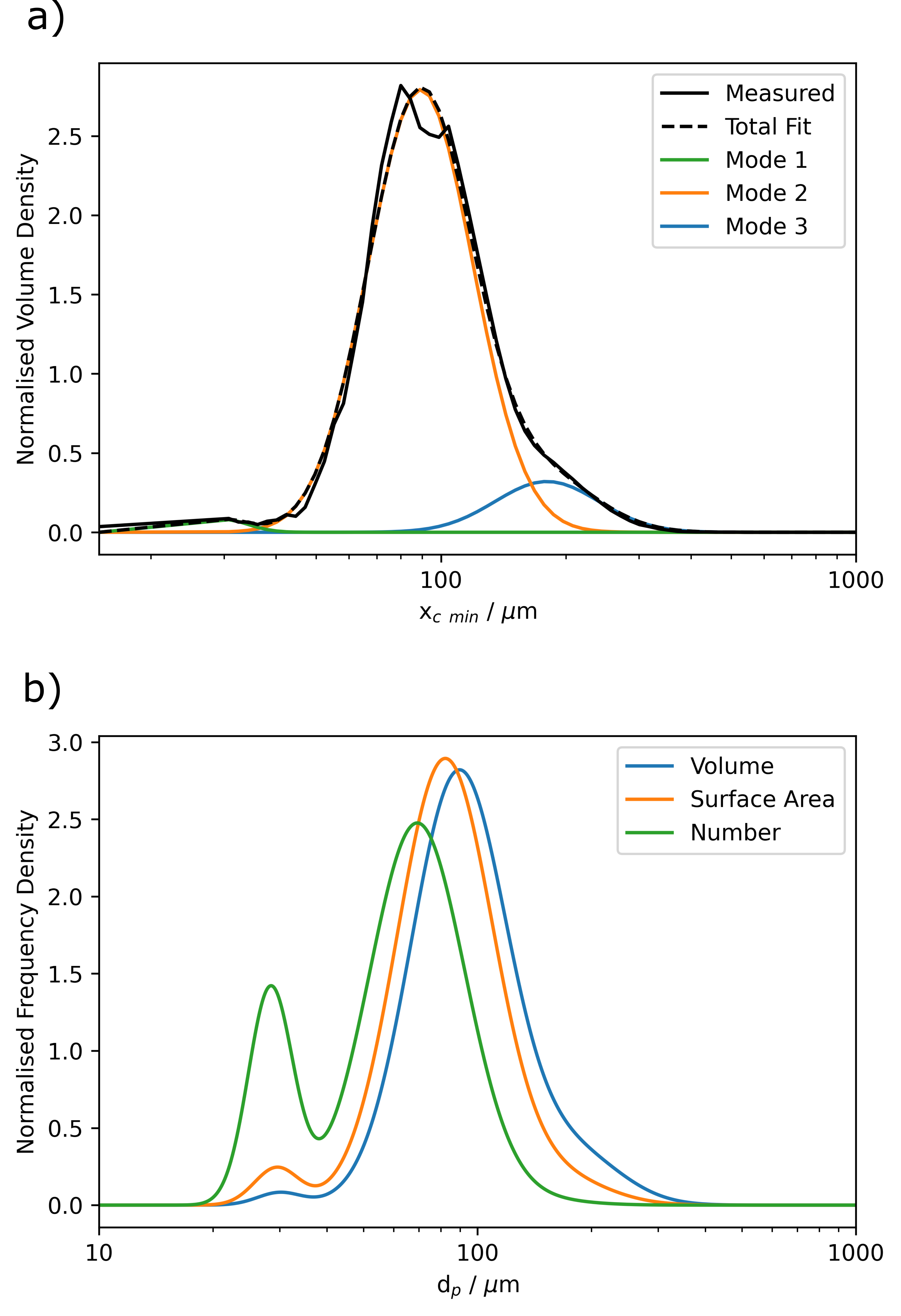}
    \caption{a) The trimodal fit for a volumetric size distribution was experimentally obtained using a CAMSIZER X2$^{\circledR}$ on a sample of ash from the Atitl\'an volcano. b) The volume distribution is converted using a spherical particle approximation to the surface area and number distributions.}
    \label{Fitted_Size_Ditsributions}
\end{figure}

The first step, as outlined in Section \ref{Model Overview}, is to find the size distribution of the particles being modelled. Naturally occurring powders tend to fit log-normally distributed modes\cite{whitby_physical_1978}. Volumetric optical size distributions for the samples were obtained from the CAMSIZER X2$^{\circledR}$ (as can be seen in Figure \ref{Fitted_Size_Ditsributions}) and the Malvern Mastersizer 3000$^{\circledR}$. The data from these were found to fit multi-modal log-normal distributions as would be expected from naturally occurring granular materials.

When converted to surface area distributions the smaller modes become more significant and are further amplified in their respective number distributions. This conversion assumes that the particles are spherical and uses the measured optical diameter, even though the aerodynamic diameter is the more relevant factor. Conversions can be made between the two quantities either utilising the particles' shape factor or empirically through measurements. In this case, no conversion was made as the relatively small adjustment (usually around 20\%\cite{chen_aerodynamic_2001}) would make an insignificant difference to the broad size distributions which span multiple orders of magnitude. The Mastersizer and Camsizer were chosen due to the operating size ranges of 10~nm to 3.5~mm and 0.8~$\mu$m to 8~mm, respectively. Multi-modal distributions with the form \begin{equation}
    f(x) = \sum_{i=1}^{n} N_i e^{\minus \frac{1}{2}\left(\frac{log_{10}(x) - log_{10}(\mu_i)}{\sigma_i}\right)}
\end{equation}
were fitted to the experimental data by minimising the total residual or maximising the R$^2$ value for the fit. For each of the $n$ log-normal modes, $\mu$ is the mean, $\sigma$ is the standard deviation in log-space, and $N$ is the modes' relative contributions. The largest modes are in the tens of microns, in line with the expectations of geological material and a much smaller mode that becomes significant when converted to number distribution can be observed at just under a micron, typical of the accumulation mode for fine geological material\cite{watson_measurement_2010}. The choice of minimising the total residual or the R$^2$ value was found to negligibly affect the fits.

\subsection{Particle dynamics model}
\label{Particle Dynamics Model}
Now that the size distributions have been found, we can look at the particle dynamics model employed to calculate the fall time for each particle to reach the cup. Similarly to Section \ref{Size Distribution Fitting}, the particles were assumed to be spherical, such that a standard set of equations could be employed and total fall time could be calculated for each particle size. To resolve the forces acting upon each particle, the following set of equations can be employed as laid out in Perry's chemical engineers' handbook\cite{perry_perrys_2008}.

\begin{figure}[t!]
    \centering
    \includegraphics[height=12cm]{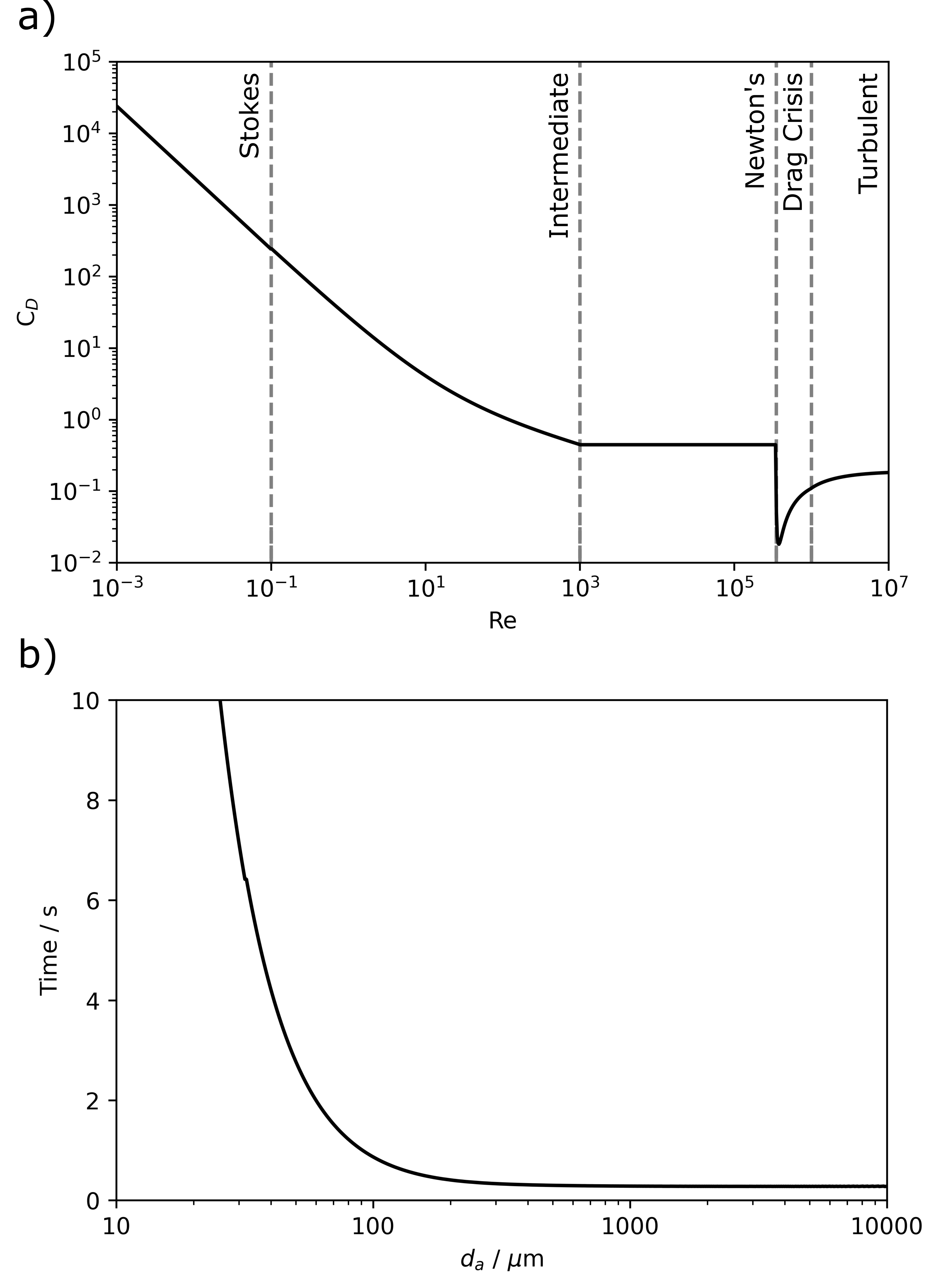}
    \caption{a) The dependency of the drag coefficient (\textrm{$C_D$}) on the Reynolds number ($Re$) with the drag regimes labelled. b) The predicted time for spherical particles to reach the Faraday cup (drop 37.25 cm) as a function of their aerodynamic diameter ($d_a$).}
    \label{Time_Distributions}
\end{figure}

For a spherical particle the drag force ($F_D$) exerted on the particle by the surrounding fluid, which is air in this case, can be expressed as
\begin{equation}
    F_D = \frac{C_{D}A_{P}\rho_f u^2}{2},
\end{equation}
where $C_D$ is the drag coefficient, $A_P$ is the cross-sectional area of the particle ($A_P = \frac{\pi d_{p}^2}{4}$ for spherical particles), $\rho_f$ is the density of the fluid, and $u$ is the relative velocity of the particle and the fluid. The drag coefficient is dependent on the Reynolds number ($Re$) in a non-simplistic way. The Reynolds number is a dimensionless quantity that represents the ratio of inertial to viscous forces and can be expressed as 
\begin{equation}
    Re = \frac{\rho_f uL}{\mu},
\end{equation}
whereby, $\mu$ is the fluid's dynamic viscosity and $L$ is the characteristic length scale, which is the particle diameter ($d_p$) in this case. In this work the fluid is air under roughly standard conditions, meaning $\rho_f$ and $\mu$ are $1.225$ kg m$^{\minus 3}$ and $1.79 \times 10^{\minus 5}$ kg m$^{\minus 1} $s$^{\minus 1}$, respectively, although both change with environmental factors such as fluctuations in temperature and humidity\cite{perry_perrys_2008, balevicius_air_2020, venczel_temperature-dependent_2021}. At low Reynolds numbers ($Re$), spherical particles will obey Stokes' law where
\begin{equation}
    C_D = \frac{24}{Re}.
\end{equation}
As the Reynolds number increases this direct relation moves into the intermediate regime ($0.1 < Re < 1,000$), where the drag coefficient relation becomes
\begin{equation}
    C_D = \left(\frac{24}{Re}\right) \left(1 + 0.14 Re^{0.70}\right).
\end{equation}
At greater still Reynolds numbers lie Newton's, drag crisis, and turbulent regimes as seen in Figure \ref{Time_Distributions}a. However, due to the scale of this work, the particles always stay within the Stokes and intermediate regimes.
The only other force that is assumed be acting on the particles is the force due to gravity ($F_g$), which naturally acts downwards with $F_g = m_pg$, where $g$ is the acceleration due to gravity and the mass of the particles $m_p$ is related to their diameter and density ($\rho_p$) by $m_p = \frac{1}{6}\pi d_p^3 \rho_p$. In this work, $\rho_p$ was assumed to be constant throughout each sample and was measured using the Anton Paar Ultrapyc 3000$^{\circledR}$ gas pycnometer.

These forces can be resolved and used with Newton's second law to find each particle's acceleration. Each time step the position and velocity of the particles can be updated, recalculating $Re$, $C_D$, and $F_D$. For numerical stability, the time step must be updated such that the Courant–Friedrichs–Lewy (CFL) condition is held at one. The CFL condition is the ratio of the simulation time step to the system's characteristic length scale. In this case, the particle can move roughly its diameter within each time step. At the very beginning, the particles start stationary, meaning the CFL condition number would be zero so a minimum time step of 0.1~ms is also implemented. Overall, this allows for the total time for each particle size to reach the cup to be determined, with the results shown in Figure \ref{Time_Distributions}b. The aerodynamic diameter here is equated to the geometric diameter, as a spherical approximation is made, although empirical measurements could be made to account for this discrepancy.

Only particles that land within the final trace time need to be calculated, as the rest will not land and hence be shown on the trace. In the case of Figure \ref{Time_Distributions}b only particles of over 20~$\mu$m would contribute to the predicted trace. Additionally, as particles approach and pass below 10~$\mu$m in size they would begin to behave diffusively and deposit on the container walls or remain dispersed in the air rather than impacting the Faraday cup\cite{ruzer_aerosols_2012}.

\subsection{Charge frequency density}
\label{Charge Frequency Density}
The next step of the modular approach requires a prediction of the separate charge contributions' dependency upon particle size. In the majority of this work the simplest cases for both the pre- and self-charging were chosen. For the pre-charging that is imparted from the loading and contact with walls, the simplest case is to assume this charging is uni-polar and is proportional to the surface area of each particle. As the particles' aerodynamic diameter is used in this model the surface area is then assumed to scale with ${d_p}^2$. Validation for this approach is provided by Section \ref{Size dependence of pre-charging}. Critically, only the shape of the distribution is assumed, so when the fitting takes place the fitted parameters may predict that the particles may be pre-charging positively or negatively and the magnitude may vary.

A slightly more involved approach is required for the self-charging (particle-particle interactions). In this work, an Event-Driven Molecular Dynamics (EDMD) hard-sphere approach was adopted, which reduces computational expense. Due to the unknown nature of tribocharging, this work does not assume the nature of the charge carrier. Therefore, a complex charge transfer mechanism, such as electron tunneling, as employed by Kok \textit{et al.} cannot be used\cite{kok_electrification_2009}. Instead, the simple case of transferring a single charge from a high-energy to a low-energy state is assumed. This can be thought of as negative (e.g. an electron) initially, although again when the fitting takes place only the shape is important and the polarity may be flipped, such that the charge carrier could be of either polarity.

\begin{figure}[b!]
    \centering
    \includegraphics[height=12cm]{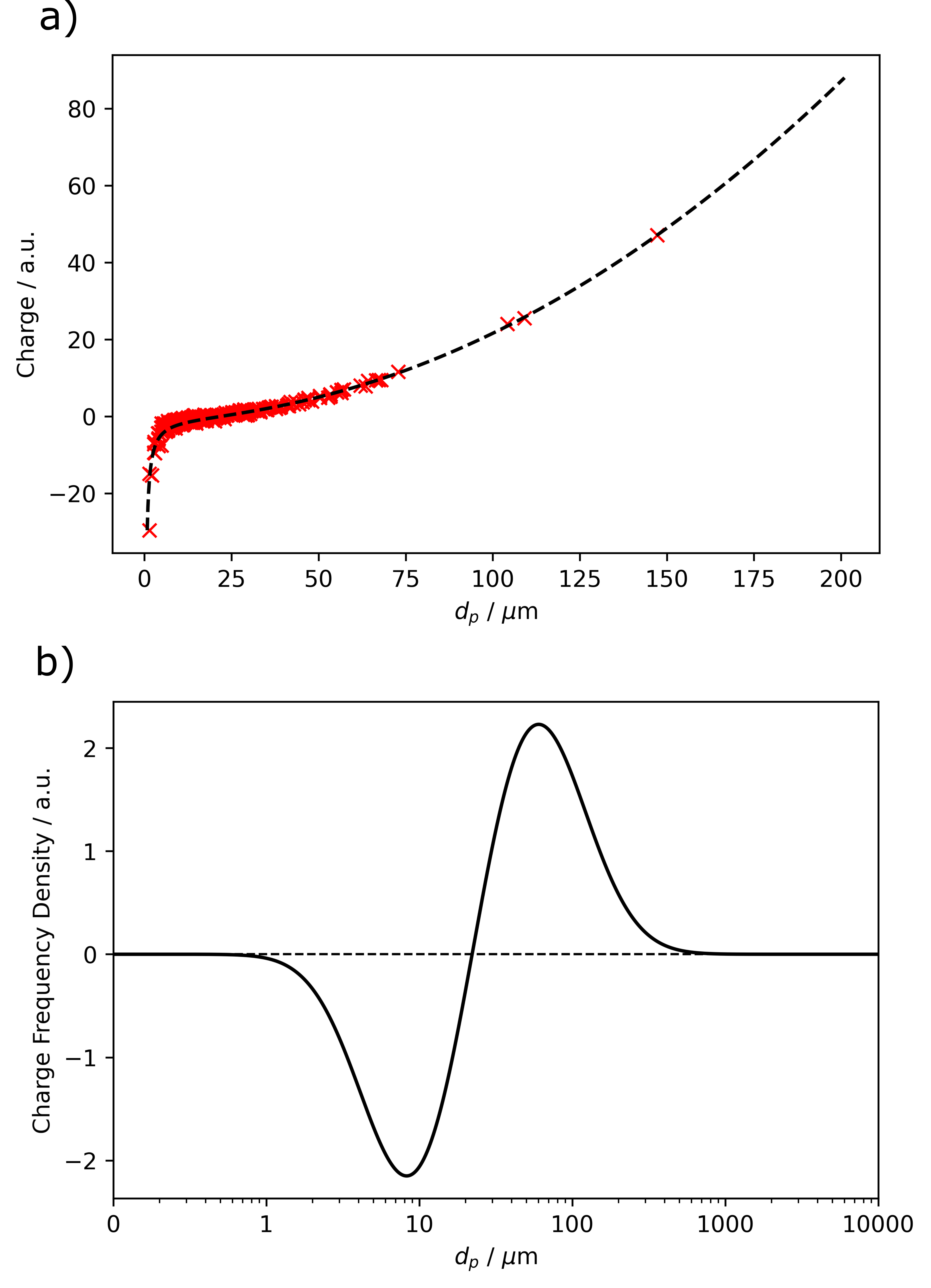}
    \caption{a) The EDMD simulated charging against size ($d_p$) for a monomodal fit of a sample of ash from Mount St. Helens where the charge carriers are negative and have arbitrary units of 1. The fit, shown as a dashed line, is a polynomial of the form $f(d_p) = a {d_p}^2 + c {d_p}^{-1}$, where $a = 0.00218$ and $c = \minus23.7$. b) The convolution of the charging and size distributions to give a charge frequency density distribution.}
    \label{Example_Charge_Ditsributions}
\end{figure}

In this EDMD model, distributions of 300 particles were used, matching the size distribution described in Section \ref{Size Distribution Fitting}. These particles were then randomly placed in a box such that their density was held constant, at a packing fraction of 0.1, by calculating the required box length for each volume of particles. If particles overlapped then one of the overlapping particles was replaced elsewhere in the box until there were no overlaps present. The time until a particle-particle collision event ($t_{col}$) between each pair of particles, such as $i$ and $j$, can be calculated from the initial distance $\bm{r}_{ij} = \bm{r}_{j}(t) - \bm{r}_{i}(t)$ and velocity $\bm{v}_{ij} = \bm{v}_{j}(t) - \bm{v}_{i}(t)$ vectors by
\begin{equation}
    t_{col} = \frac{-b - \sqrt{b^2-\bm{v}_{ij}^2\left(\bm{r}_{ij}^2 - \sigma_{ij}^2\right)}}{\bm{v}_{ij}^2},
\end{equation}
where $b = \bm{r}_{ij} \cdot \bm{v}_{ij}$ and $\sigma_{ij}$ is the sum of the particles' radii. 

After an elastic collision, the change in the particles' velocity was calculated using the particles' respective masses ($m_i$ and $m_j$), their velocities and their unit vector of separation $\bm{\hat{r}}_{ij}$. By first calculating the effective mass $m_{eff}$ and impact speed $v_{imp}$,
\begin{equation}
    m_{eff} = \frac{m_im_j}{m_i+m_j} \qquad \textrm{and} \qquad v_{imp} = \bm{\hat{r}}_{ij} \cdot \left(\bm{v}_i-\bm{v}_j\right),
\end{equation}
the impulse magnitude was determined by
\begin{equation}
    J = \left(1+\epsilon\right)m_{eff}v_{imp},
\end{equation}
where $\epsilon$ is the coefficient of restitution and is equal to 1 for an elastic collision. Thus the change in velocity is calculated as
\begin{equation}
    \delta \bm{v}_{i} = - \frac{J}{m_i}\bm{\hat{r}}_{ij}  \qquad \textrm{and} \qquad \delta \bm{v}_{j} = \frac{J}{m_i}\bm{\hat{r}}_{ij}.
\end{equation}

A hard wall treatment was used in this case for simplicity, meaning if a particle collided with a wall its velocity vector in the direction of the wall would be inverted, although a periodic boundary condition could also be used\cite{smallenburg_efficient_2022}.

Initially, these particles were assigned random velocities before being allowed to equilibriate for a few thousand steps, such that they could reach a thermodynamic equilibrium (defined as their kinetic energies fitting a Maxwell-Boltzmann distribution) before the charge transfer mechanism was turned on. The final arbitrary charges were then fit to a polynomial of the form $f(d_p) = a {d_p}^b + c {d_p}^{d} + e$, where d$_p$ is the particle diameter and $a$, $b$, $c$, $d$, and $e$ are fitting parameters, such that $a > 0$ and $c < 0$. For the positive term it was found that $b$ was usually around 2 as at large diameters the surface area ($\propto$ ${d_p}^2$) dominates the fitting. For the negative term $d$ was usually around $-1$, which is close to what may be expected from the increase collisions of smaller particles due to their greater velocity at the same kinetic energy ($\propto$ ${d_p}^{-\frac{3}{2}}$) but slightly offset by the smaller collision diameter reducing the effective collision cross-section. The fit with constraints $b = 2$, $d = -1$, and $e = 0$ shows very good agreement with the data, giving an $R^2$ value of 0.996 in Figure \ref{Example_Charge_Ditsributions}a. Overall, this fit is in line with expectations where the small particles are negatively charged and large particles are positively charged for a negative charge carrier, such as an electron\cite{lacks_effect_2007}.

As the size distribution is represented as frequency density and the charge distribution shape is known as a function of size, the two can be multiplied to give a charge frequency density distribution, demonstrated in Figure \ref{Example_Charge_Ditsributions}b. For self-charging, there should naturally be no net charge across the whole distribution due to the conservation of charge, meaning the total integral of the charge frequency density distribution should be zero. In reality, the distributions integrate to near zero but not exactly due to the discretisation of the particle sizes used in the EDMD simulation. If an infinite number of particles were used the integral of the arising charge frequency density distributions of self-charging should limit to zero. In the case of Mount St. Helen's, the total integration deviates from zero by 2.2\% as a fraction of the functions' total variation. This process was then repeated for expected charge distribution arising from pre-charging sources ($Q$ $\propto$ ${d_p}^2$) to obtain the respective pre-charging frequency density distribution.

\subsection{Overall trace and fit}
Finally, the charge frequency densities from Section \ref{Charge Frequency Density} can be combined with the drop time function from Section \ref{Particle Dynamics Model} to get the overall charge trace expected over time. This is done by checking, at each time step, which particles are predicted to have landed and integrating the charge frequency density from infinity to that point. These total charging traces were then fit to the experimentally obtained charging traces, similarly to those laid out in section \ref{Size Distribution Fitting}, minimising the $R^2$ values of the fits by solving for the ratios between the pre- and self-charging with a Nelder-Mead algorithm utilising the \href{https://docs.scipy.org/doc/scipy/reference/generated/scipy.optimize.minimize.html}{scipy.optimize.minimize} package for \href{https://docs.python.org/3/}{Python 3.12.2}. To quantify the pre- to self-charging ratio, the total integral of both charge frequency density distributions was taken and then multiplied by the relative amount of pre- to self-charging required for the predicted trace fitted.

This approach makes the assumption that the pre- and self-charging do not interact. However, due to the modular nature of this approach, more sophisticated models in both the particle dynamics and charging distribution stages could attempt to account for such interactions.

\section{Results and discussion}
\label{Results and Discussion}
\subsection{Volcanic ash}

\begin{figure}[b!]
    \centering
    \includegraphics[height=12cm]{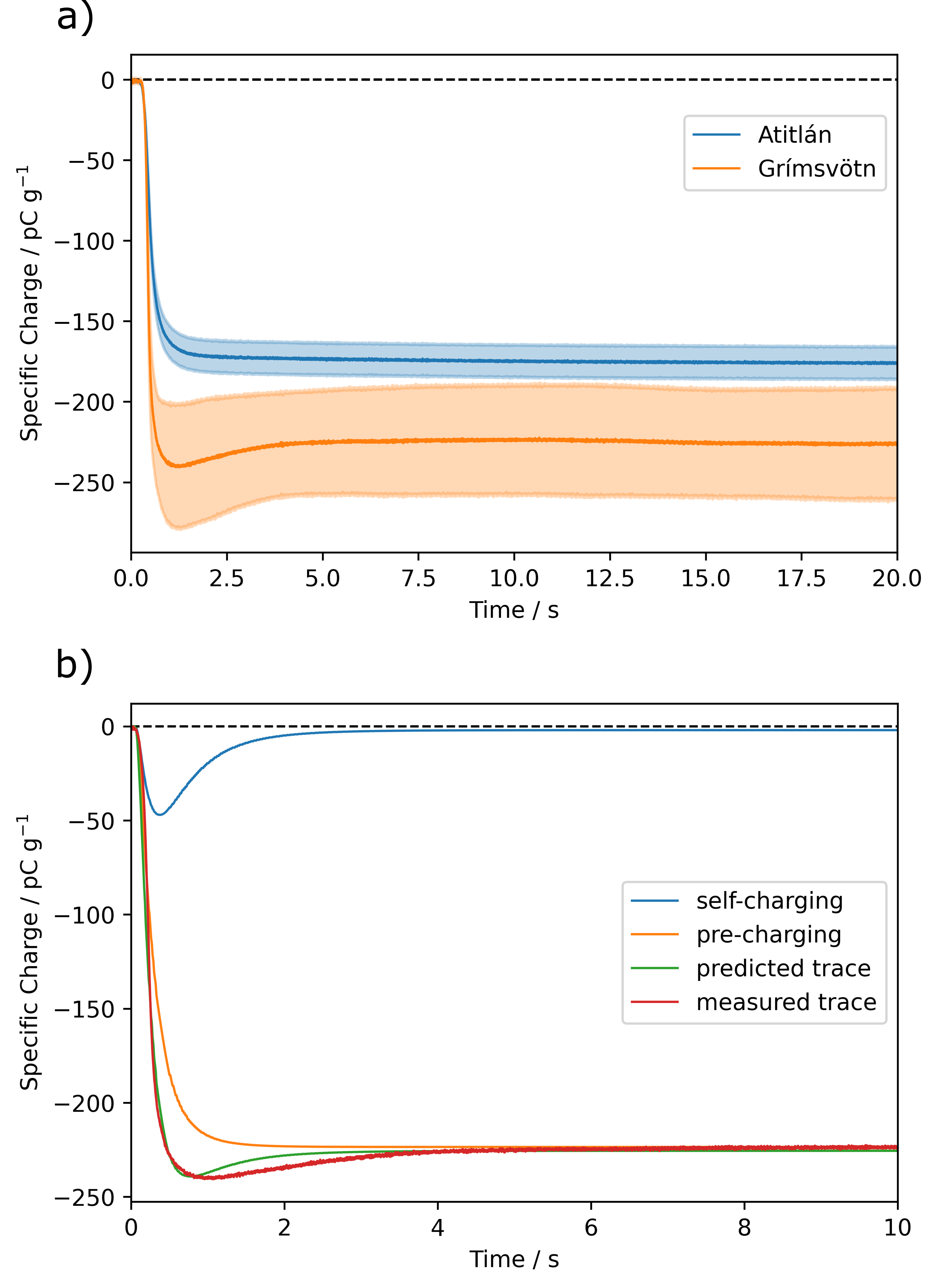}
    \caption{a) The averaged Faraday cup traces for volcanic ash from Atitl\'an and Gr\'imsv\"otn. The standard error on the mean at each timestep is shown by their respective coloured areas. b) The fitted predicted trace for the Faraday cup plotted alongside the measured (experimental) trace for Gr\'imsv\"otn with the separated self- and pre-charging components shown (R$^2$ = 0.975).}
    \label{Example_Trace_Fit}
\end{figure}

Now that we have a method to separate the charging contributions, the powder charging for various application cases can be investigated. Volcanic ash was used due to the availability of suitable samples, and the role of near-vent triboelectrification in volcanic lightning. The previously outlined methodology was followed, taking measured traces and size distributions of volcanic ash samples, to produce fitted charging components as seen in Figure \ref{Example_Trace_Fit}.

For each ash sample, the smallest particle fraction ($d_p < 63~\mu$m) was removed. This because the smallest particles are harder to model, due to some falling in the diffuse regime so will stick to walls and form agglomerates. Some samples such as Gr\'imsv\"otn ash also exhibited decay after being dropped due to the conductivity of air. This can be accounted for by adding back the expected exponential decay at each time step in the measured trace to achieve a trace without decay. The conductivity of air in the lab has previously been shown to be $2.85 \times 10^{\minus 14}$ Sm$^{-1}$, yielding a time constant of $310.5$ s for the decay\cite{reid_lab-based_2024}. This decay is only observed in certain materials, possibly due to the particles' electrical conductivity or surface chemistry playing a role in the decay of the static charge. For the fits seen in Figure \ref{Example_Trace_Fit} the $R^2$ value is minimised, prioritising the immediate drop where there is more deviation from the fit. However, a minimisation of the total residuals, which prioritises fitting the later tail of the trace can also helpful in indicating the extent of pre- and self-charging.

From the charge trace fits, we can now determine the extent of self- and pre-charging contributions. One reasonable approach to assigning an overall value to the ratio of these components' relative contributions is to take the ratio of the integrals of the respective absolute size frequency density distributions. For Gr\'imsv\"otn ash the ratio of pre-:self-charging is 2.7:1, indicating the two types of charging have initially the same polarity with 27\% of the charging originating from particle-particle interactions. For Atitl\'an, this pre-charging only makes up 7\% of the total charging. Overall, the greater self-charging indicates that the ash from Gr\'imsv\"otn is expected to charge more from particle-particle interactions in a less-bounded system such as a volcanic vent. However, the total charging (including pre-charging) is still significant in some cases as it may give information on the saturation charge of the sample. The larger degree of self-charging for Gr\'imsv\"otn ash is not inconsistent with observations of many registered lightning events per day upon eruption of the Gr\'imsv\"otn volcano\cite{arason_charge_2011}. However, it is hard to make a direct comparison between laboratory charging investigations and observed lightning events due to the many other variables and charging mechanisms involved, such as fractoelectrification and ``dirty thunderstorms''\cite{cimarelli_volcanic_2022, mcnutt_volcanic_2010}.


\subsection{Validation case}
\label{Validation Case}

\begin{figure}[t!]
    \centering
    \includegraphics[height=12cm]{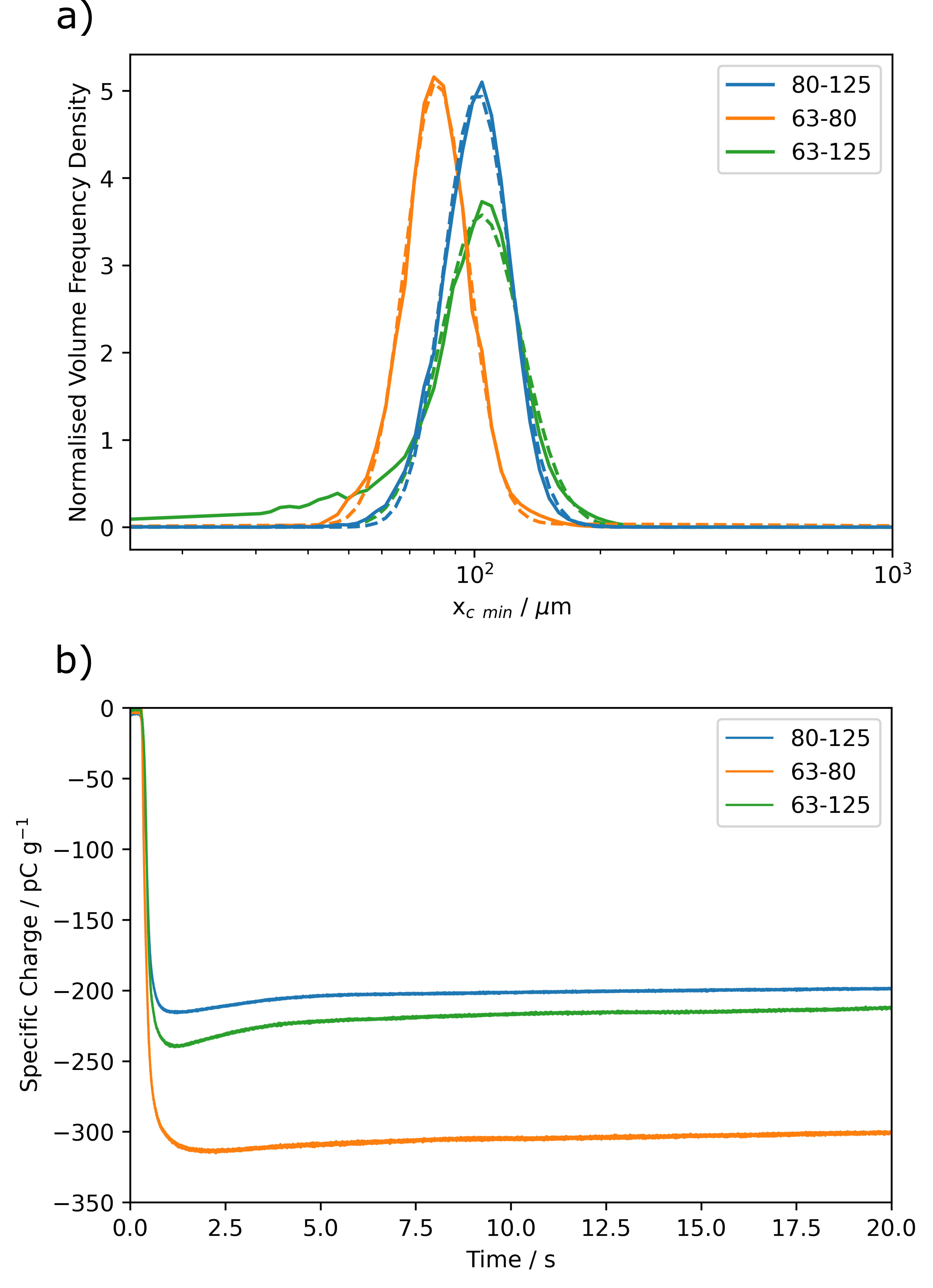}
    \caption{a) The volumetric size distributions for sieved fractions of Gr\'imsv\"otn ash, obtained using a CAMSIZER X2$^{\circledR}$ and fit with monomodal log-normal distributions. The ranges in the legend of the sieved fractions have the units of microns. b) The Faraday cup traces for the corresponding sieved fractions of Gr\'imsv\"otn ash.}
    \label{G90}
\end{figure}

To validate the new Faraday cup modular analysis approach a case with an expected outcome has to be chosen to test the preliminary results against. One expected outcome is that broader size distributions would be expected to exhibit more self-charging due to the greater asymmetry between particle sizes. Similar amounts of pre-charging and self-charging would also be preferable in order to show greater variation in charging ratios between the different samples. For this purpose, sieved fractions of Gr\'imsv\"otn ash with differing breadths of size distribution were sieved and analysed. The size distributions of the sieved fractions and their Faraday cup traces can be seen in Figure \ref{G90}.

By applying the modular analysis approach for the broad sieving fraction (63 $<$ d$_p$ $<$ 125 $\mu$m) of Gr\'imsv\"otn ash, as seen in Figure \ref{G90}c, the pre:self-charging ratio was found to be 2.4:1. This indicates that nearly a third of the charging originated from self-charging. For the slightly narrower fraction (80 $<$ d$_p$ $<$ 125 $\mu$m) the ratio was 2.9:1 indicating around a quarter of the charging was from self-charging. This trend continued to the much narrower fraction of (80 $<$ d$_p$ $<$ 125 $\mu$m) with a ratio of 1:4.8 indicating only around a sixth of the charging is arising from self-charging. The analysis demonstrates that the broader size fractions show both a greater proportion and magnitude of their electrification arising from particle-particle interactions. This is expected based on the increased asymmetry in particle size, providing evidence for the validity of the new analysis.

\subsection{Size dependence of pre-charging}
\label{Size dependence of pre-charging}
Utilising the charging interpretation methodology outlined in the previous section, different charging components can now be investigated in isolation. The size dependence of these charging processes is of great importance to both natural and industrial applications, for example with regards to ignition risks of chemical processes which may be carried out over a variety of operational scales and particle sizes. To begin with, powered samples of labradorite were chosen to investigate the scaling of pre-charging - it was found to charge with almost exclusively pre-charging in this setup. Labradorite is broadly relevant to the tribocharging of volcanic ash as X-Ray Diffraction (XRD) analysis found a reasonable match between labradorite and ash from Volc\'{a}n de Fuego\cite{ohara_faraday_2024}. Samples were sieved to change their size distributions and then dropped to get their charging traces.

\begin{figure}[t!]
    \centering
    \includegraphics[height=5.75cm]{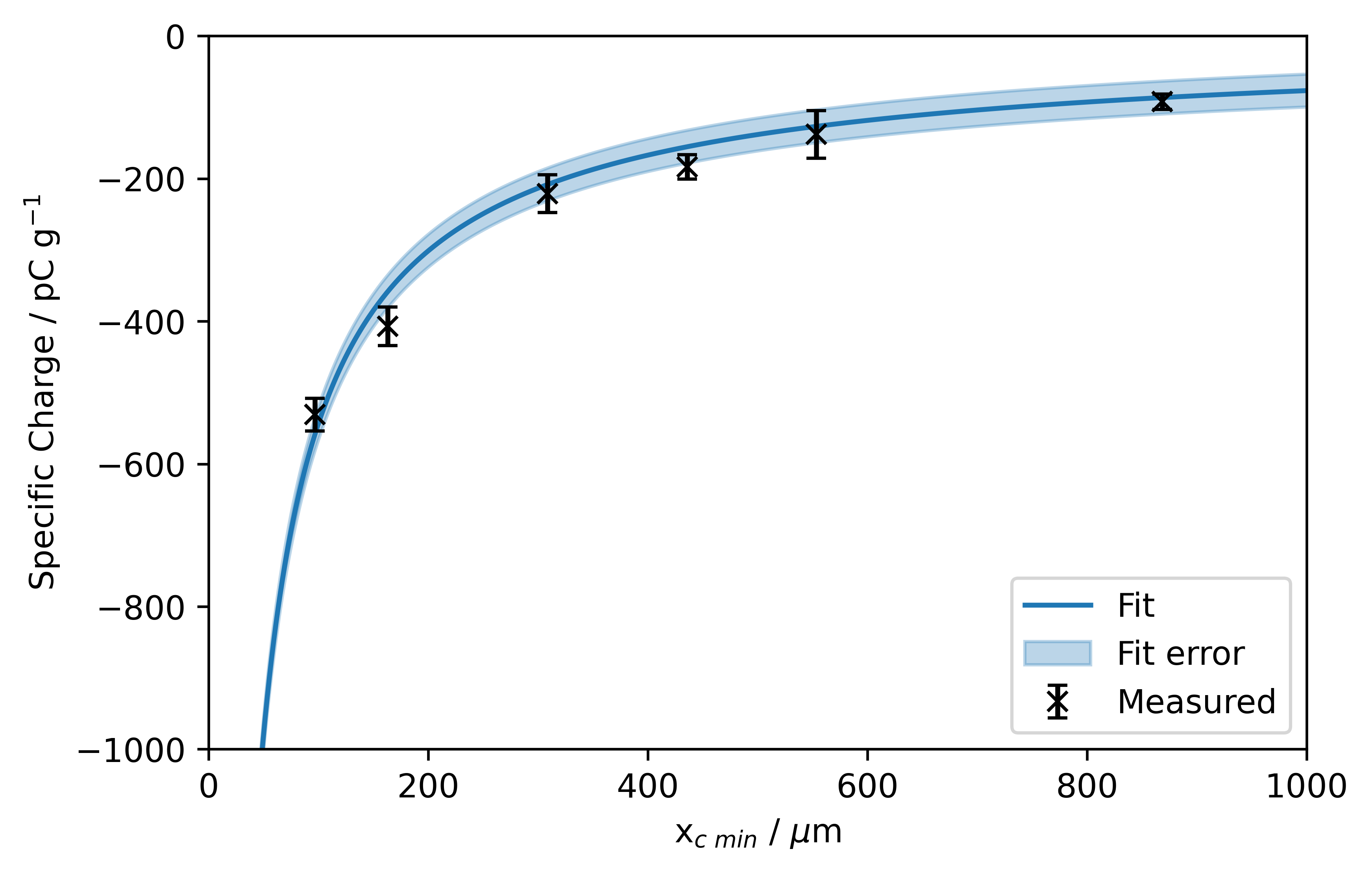}
    \caption{The average specific charge drop for sieved fractions of powdered labradorite plotted against the average sieving diameter (x$_{c\ min}$)$^{\ast}$. The fit is a power-law function of the form $Q = a$ $x_{c\ min}^b$, where $a = \minus 27000 \pm 5000$ and $b = \minus~0.85~\pm~0.03$, respectively. The error bars represent one standard error on the mean in the specific charge drops for each size range and the fit error represents the area within one standard deviation of the residuals from the fit.}
    \label{Labradorite_Fits}
\end{figure}

The measured labradorite self-charging was found to fit a power-law function as shown in Figure \ref{Labradorite_Fits} with an $R^2$ value of 0.98, with a power of $\minus~0.85~\pm~0.03$. This is likely due to the increased contact area between the particles and the walls of the container or any other handling apparatus. To approximate this effect we could assume the container walls are effectively flat and the particles are mono-disperse spheres of radius $r$. Then the approximate amount of each particles surface area that is within a distance ($L$) of the surface is $2\pi rL$. If the spheres are close packing hexagonally in 2D against an area of the flat surface, then the packing fraction is constant at  $\pi / 2\sqrt{3}$ giving $1 / 2r^2\sqrt{3}$ particles per area of the flat surface ($A$). Therefore, the total effective area ($A_{eff}$) can be approximated by: \begin{equation} A_{eff} = \frac{\pi AL}{r\sqrt{3}}. \end{equation} The effective surface area is thus expected to scale inversely proportionally to particle size. The experimental result of $\minus~0.85~\pm~0.03$ is not far from this approximation, which predicts $\minus 1$. The discrepancy most likely arises from the real material being poly-disperse and non-spherical.

This trend explains why the smallest size fraction in Section \ref{Validation Case} displayed the greatest overall magnitude of charging despite having a smaller self-charging component. More investigation is needed to include other variables such as accounting for differently shaped particles to probe the relation to contact area and the dependence of the self-charging on average particle size.

\footnotetext{* x$_{c\ min}$ is calculated from the shortest chord out of the measured set of maxima for a particle projection in the Camsizer X2 and is used here as a proxy for sieving diameter. The average referred to here is of the log-normal fit hence is the mean, median, and mode.} 

\section{Conclusions}
\label{Conclusions}

This work has developed a novel modular approach to understanding the triboelectric charging contributions from different sources in the analysis of Faraday cup measurements of insulating granular material. This allowed for the comparison of ash from the Gr\'imsv\"otn and Atitl\'an volcanoes, which found that 27\% of the charging for Gr\'imsv\"otn ash came from particle-particle interactions but only 7\% for Atitl\'an. Analysis of sieved Gr\'imsv\"otn ash validated the modular approach as broader size fractions were found to show a greater proportion and magnitude of particle-particle charging, as would be expected from the increased particle asymmetry. Our findings suggest that in cases where some samples may have appeared to charge more than others in previous analyses, the observed charging may have been primarily due to pre-charging. As a result, these samples may exhibit reduced charging in unbounded systems such as volcanic plumes or dust devils. Conversely, samples that exhibit more self-charging, rather than pre-charging, are likely to exhibit a greater degree of charge build-up in such environments. Finally, for simplified samples of powdered labradorite the pre-charging was found to have an inverse relation with average particle size, indicating the contact area is significant in this charging contribution. Future work could investigate the scale-up of self-charging triboelectrification and the dependency of particle shape on the charging behaviour.

The outlined approach is broadly generalisable to any insulating granular material, although the experimental setup outlined in this work operates best for materials with average particle diameters ($d_p$) in the range 20~$\mu$m < $d_p$ < 200~$\mu$m (for particles with a density of around $1.5$ to $3.0$~g cm$^{\minus 3}$). This operating range could be altered by utilising a different drop height, whereby a taller experimental setup would allow for particles of greater diameter or lower density to be distinguished. The modular nature of the approach also allows for fine-tuning to each application case. Overall, this novel interpretation methodology will allow for greater insights into the granular triboelectric charging behaviour for a host of natural and industrial applications.

\section*{Author contributions}
Tom O'Hara: conceptualisation, methodology, software, formal analysis, validation, data curation, writing – original draft, writing – review and editing, project administration. David Reid: software, writing – review and editing. Gregory Marsden: investigation, writing – review and editing. Karen Aplin: supervision, resources, writing – review and editing, funding acquisition.

\section*{Conflicts of interest}
There are no conflicts to declare.

\section*{Data availability}
All the data for this article, including Faraday cup traces, size distributions, simulation outputs, and density measurements are available through the Materials Data Facility\cite{ohara_faraday_2024}. The full code for the EDMD simulation and data processing are available under a GPL-3.0 license\cite{ohara_vb22224drop_modelling_and_fitting_2024, ohara_vb22224simplified_edmd_triboelectric_2024}.

\section*{Acknowledgements}
The authors acknowledge funding from the Engineering and Physical Sciences Research Council through the Centre for Doctoral Training in Aerosol Science (no. EP/S023593/1) and UK Space Agency Aurora program (no. ST/W002485/1). We extend our thanks to Rowan Dejardin from the School of Geographical Sciences and Hayley Goodes from the School of Earth Sciences (University of Bristol) for their assistance with access to equipment. We would also like to thank \TH{o}r\dh ur Arason and Krist\'in Hermanns\dh\'ottir (Icelandic Meteorological Office) as well as Matthew Watson and Alec Bennett (University of Bristol) for providing the volcanic ash samples used.



\balance


\bibliography{main} 

\providecommand*{\mcitethebibliography}{\thebibliography}
\csname @ifundefined\endcsname{endmcitethebibliography}
{\let\endmcitethebibliography\endthebibliography}{}
\begin{mcitethebibliography}{43}
\providecommand*{\natexlab}[1]{#1}
\providecommand*{\mciteSetBstSublistMode}[1]{}
\providecommand*{\mciteSetBstMaxWidthForm}[2]{}
\providecommand*{\mciteBstWouldAddEndPuncttrue}
  {\def\EndOfBibitem{\unskip.}}
\providecommand*{\mciteBstWouldAddEndPunctfalse}
  {\let\EndOfBibitem\relax}
\providecommand*{\mciteSetBstMidEndSepPunct}[3]{}
\providecommand*{\mciteSetBstSublistLabelBeginEnd}[3]{}
\providecommand*{\EndOfBibitem}{}
\mciteSetBstSublistMode{f}
\mciteSetBstMaxWidthForm{subitem}
{(\emph{\alph{mcitesubitemcount}})}
\mciteSetBstSublistLabelBeginEnd{\mcitemaxwidthsubitemform\space}
{\relax}{\relax}

\bibitem[Houghton \emph{et~al.}(2013)Houghton, Aplin, and Nicoll]{houghton_triboelectric_2013}
I.~M.~P. Houghton, K.~L. Aplin and K.~A. Nicoll, \emph{Physical Review Letters}, 2013, \textbf{111}, 118501\relax
\mciteBstWouldAddEndPuncttrue
\mciteSetBstMidEndSepPunct{\mcitedefaultmidpunct}
{\mcitedefaultendpunct}{\mcitedefaultseppunct}\relax
\EndOfBibitem
\bibitem[Reid and Aplin(2024)]{reid_lab-based_2024}
D.~P. Reid and K.~L. Aplin, \emph{Journal of Physics: Conference Series}, 2024, \textbf{2702}, 012020\relax
\mciteBstWouldAddEndPuncttrue
\mciteSetBstMidEndSepPunct{\mcitedefaultmidpunct}
{\mcitedefaultendpunct}{\mcitedefaultseppunct}\relax
\EndOfBibitem
\bibitem[Harrison \emph{et~al.}(2008)Harrison, Aplin, Leblanc, and Yair]{harrison_planetary_2008}
R.~G. Harrison, K.~L. Aplin, F.~Leblanc and Y.~Yair, \emph{Space Science Reviews}, 2008, \textbf{137}, 5--10\relax
\mciteBstWouldAddEndPuncttrue
\mciteSetBstMidEndSepPunct{\mcitedefaultmidpunct}
{\mcitedefaultendpunct}{\mcitedefaultseppunct}\relax
\EndOfBibitem
\bibitem[Rayborn and Jellinek(2022)]{rayborn_random_2022}
L.~Rayborn and A.~M. Jellinek, \emph{Journal of Geophysical Research: Solid Earth}, 2022, \textbf{127}, e2021JB023599\relax
\mciteBstWouldAddEndPuncttrue
\mciteSetBstMidEndSepPunct{\mcitedefaultmidpunct}
{\mcitedefaultendpunct}{\mcitedefaultseppunct}\relax
\EndOfBibitem
\bibitem[Cimarelli \emph{et~al.}(2022)Cimarelli, Behnke, Genareau, Harper, and Van~Eaton]{cimarelli_volcanic_2022}
C.~Cimarelli, S.~Behnke, K.~Genareau, J.~M. Harper and A.~R. Van~Eaton, \emph{Bulletin of Volcanology}, 2022, \textbf{84}, 78\relax
\mciteBstWouldAddEndPuncttrue
\mciteSetBstMidEndSepPunct{\mcitedefaultmidpunct}
{\mcitedefaultendpunct}{\mcitedefaultseppunct}\relax
\EndOfBibitem
\bibitem[{NFPA}(2019)]{nfpa_recommended_2019}
{NFPA}, \emph{Recommended {Practice} on {Static} {Electricity}}, NFPA, 2019\relax
\mciteBstWouldAddEndPuncttrue
\mciteSetBstMidEndSepPunct{\mcitedefaultmidpunct}
{\mcitedefaultendpunct}{\mcitedefaultseppunct}\relax
\EndOfBibitem
\bibitem[Beretta \emph{et~al.}(2020)Beretta, Hörmann, Hainz, Hsiao, and Paudel]{beretta_investigation_2020}
M.~Beretta, T.~Hörmann, P.~Hainz, W.~Hsiao and A.~Paudel, \emph{International Journal of Pharmaceutics}, 2020, \textbf{591}, 120015\relax
\mciteBstWouldAddEndPuncttrue
\mciteSetBstMidEndSepPunct{\mcitedefaultmidpunct}
{\mcitedefaultendpunct}{\mcitedefaultseppunct}\relax
\EndOfBibitem
\bibitem[Lacks and Shinbrot(2019)]{lacks_long-standing_2019}
D.~J. Lacks and T.~Shinbrot, \emph{Nature Reviews Chemistry}, 2019, \textbf{3}, 465--476\relax
\mciteBstWouldAddEndPuncttrue
\mciteSetBstMidEndSepPunct{\mcitedefaultmidpunct}
{\mcitedefaultendpunct}{\mcitedefaultseppunct}\relax
\EndOfBibitem
\bibitem[Cezan \emph{et~al.}(2019)Cezan, Nalbant, Buyuktemiz, Dede, Baytekin, and Baytekin]{cezan_control_2019}
S.~D. Cezan, A.~A. Nalbant, M.~Buyuktemiz, Y.~Dede, H.~T. Baytekin and B.~Baytekin, \emph{Nature Communications}, 2019, \textbf{10}, 276\relax
\mciteBstWouldAddEndPuncttrue
\mciteSetBstMidEndSepPunct{\mcitedefaultmidpunct}
{\mcitedefaultendpunct}{\mcitedefaultseppunct}\relax
\EndOfBibitem
\bibitem[Wilcke(1757)]{wilcke_dispvtatio_1757}
J.~C. Wilcke, \emph{Dispvtatio {Physica} {Experimentalis}, {De} {Electricitatibvs} {Contrariis}: {Qvam} {Consentiente}... {In} {Academia} {Rostochiensi}, {Ad} {D}. {XIII}. {Octobr}. {A}. {MDCCLVII}.... {Dissertation}}, 1757\relax
\mciteBstWouldAddEndPuncttrue
\mciteSetBstMidEndSepPunct{\mcitedefaultmidpunct}
{\mcitedefaultendpunct}{\mcitedefaultseppunct}\relax
\EndOfBibitem
\bibitem[Xu and Grosshans(2023)]{xu_experimental_2023}
W.~Xu and H.~Grosshans, \emph{Journal of Loss Prevention in the Process Industries}, 2023, \textbf{81}, 104970\relax
\mciteBstWouldAddEndPuncttrue
\mciteSetBstMidEndSepPunct{\mcitedefaultmidpunct}
{\mcitedefaultendpunct}{\mcitedefaultseppunct}\relax
\EndOfBibitem
\bibitem[Jallo and Dave(2015)]{jallo_explaining_2015}
L.~J. Jallo and R.~N. Dave, \emph{Journal of Pharmaceutical Sciences}, 2015, \textbf{104}, 2225--2232\relax
\mciteBstWouldAddEndPuncttrue
\mciteSetBstMidEndSepPunct{\mcitedefaultmidpunct}
{\mcitedefaultendpunct}{\mcitedefaultseppunct}\relax
\EndOfBibitem
\bibitem[Biegaj \emph{et~al.}(2017)Biegaj, Rowland, Lukas, and Heng]{biegaj_surface_2017}
K.~W. Biegaj, M.~G. Rowland, T.~M. Lukas and J.~Y.~Y. Heng, \emph{ACS Omega}, 2017, \textbf{2}, 1576--1582\relax
\mciteBstWouldAddEndPuncttrue
\mciteSetBstMidEndSepPunct{\mcitedefaultmidpunct}
{\mcitedefaultendpunct}{\mcitedefaultseppunct}\relax
\EndOfBibitem
\bibitem[Cruise \emph{et~al.}(2022)Cruise, Hadler, Starr, and Cilliers]{cruise_effect_2022}
R.~D. Cruise, K.~Hadler, S.~O. Starr and J.~J. Cilliers, \emph{Journal of Physics D: Applied Physics}, 2022, \textbf{55}, 185306\relax
\mciteBstWouldAddEndPuncttrue
\mciteSetBstMidEndSepPunct{\mcitedefaultmidpunct}
{\mcitedefaultendpunct}{\mcitedefaultseppunct}\relax
\EndOfBibitem
\bibitem[Peart(2001)]{peart_powder_2001}
J.~Peart, \emph{KONA Powder and Particle Journal}, 2001, \textbf{19}, 34--45\relax
\mciteBstWouldAddEndPuncttrue
\mciteSetBstMidEndSepPunct{\mcitedefaultmidpunct}
{\mcitedefaultendpunct}{\mcitedefaultseppunct}\relax
\EndOfBibitem
\bibitem[Cruise \emph{et~al.}(2023)Cruise, Starr, Hadler, and Cilliers]{cruise_triboelectric_2023}
R.~D. Cruise, S.~O. Starr, K.~Hadler and J.~J. Cilliers, \emph{Scientific Reports}, 2023, \textbf{13}, 15178\relax
\mciteBstWouldAddEndPuncttrue
\mciteSetBstMidEndSepPunct{\mcitedefaultmidpunct}
{\mcitedefaultendpunct}{\mcitedefaultseppunct}\relax
\EndOfBibitem
\bibitem[Waitukaitis \emph{et~al.}(2014)Waitukaitis, Lee, Pierson, Forman, and Jaeger]{waitukaitis_size-dependent_2014}
S.~R. Waitukaitis, V.~Lee, J.~M. Pierson, S.~L. Forman and H.~M. Jaeger, \emph{Physical Review Letters}, 2014, \textbf{112}, 218001\relax
\mciteBstWouldAddEndPuncttrue
\mciteSetBstMidEndSepPunct{\mcitedefaultmidpunct}
{\mcitedefaultendpunct}{\mcitedefaultseppunct}\relax
\EndOfBibitem
\bibitem[Zhao \emph{et~al.}(2002)Zhao, Castle, and Inculet]{zhao_measurement_2002}
H.~Zhao, G.~Castle and I.~Inculet, \emph{Journal of Electrostatics}, 2002, \textbf{55}, 261--278\relax
\mciteBstWouldAddEndPuncttrue
\mciteSetBstMidEndSepPunct{\mcitedefaultmidpunct}
{\mcitedefaultendpunct}{\mcitedefaultseppunct}\relax
\EndOfBibitem
\bibitem[Yeo \emph{et~al.}(2023)Yeo, Wang, Dove, and Horányi]{yeo_laboratory_2023}
L.~H. Yeo, X.~Wang, A.~Dove and M.~Horányi, \emph{Advances in Space Research}, 2023, \textbf{72}, 1861--1869\relax
\mciteBstWouldAddEndPuncttrue
\mciteSetBstMidEndSepPunct{\mcitedefaultmidpunct}
{\mcitedefaultendpunct}{\mcitedefaultseppunct}\relax
\EndOfBibitem
\bibitem[Thomas \emph{et~al.}(2008)Thomas, Saleh, Guigon, and Czechowski]{thomas_tribocharging_2008}
A.~Thomas, K.~Saleh, P.~Guigon and C.~Czechowski, \emph{Journal of Physics: Conference Series}, 2008, \textbf{142}, 012031\relax
\mciteBstWouldAddEndPuncttrue
\mciteSetBstMidEndSepPunct{\mcitedefaultmidpunct}
{\mcitedefaultendpunct}{\mcitedefaultseppunct}\relax
\EndOfBibitem
\bibitem[Maxwell(1873)]{maxwell_treatise_1873}
J.~C. Maxwell, \emph{Clarendon Press google schola}, 1873, \textbf{2}, 3408--3425\relax
\mciteBstWouldAddEndPuncttrue
\mciteSetBstMidEndSepPunct{\mcitedefaultmidpunct}
{\mcitedefaultendpunct}{\mcitedefaultseppunct}\relax
\EndOfBibitem
\bibitem[Kucerovsky and Kucerovsky(2003)]{kucerovsky_analysis_2003}
D.~Kucerovsky and Z.~Kucerovsky, \emph{Journal of Physics D: Applied Physics}, 2003, \textbf{36}, 2407--2416\relax
\mciteBstWouldAddEndPuncttrue
\mciteSetBstMidEndSepPunct{\mcitedefaultmidpunct}
{\mcitedefaultendpunct}{\mcitedefaultseppunct}\relax
\EndOfBibitem
\bibitem[Sosolik \emph{et~al.}(2000)Sosolik, Lavery, Dahl, and Cooper]{sosolik_technique_2000}
C.~E. Sosolik, A.~C. Lavery, E.~B. Dahl and B.~H. Cooper, \emph{Review of Scientific Instruments}, 2000, \textbf{71}, 3326--3330\relax
\mciteBstWouldAddEndPuncttrue
\mciteSetBstMidEndSepPunct{\mcitedefaultmidpunct}
{\mcitedefaultendpunct}{\mcitedefaultseppunct}\relax
\EndOfBibitem
\bibitem[Sunil \emph{et~al.}(2023)Sunil, Shukla, and Sharma]{sunil_study_2023}
K.~Sunil, R.~Shukla and A.~Sharma, \emph{Plasma Science and Technology}, 2023,  045402\relax
\mciteBstWouldAddEndPuncttrue
\mciteSetBstMidEndSepPunct{\mcitedefaultmidpunct}
{\mcitedefaultendpunct}{\mcitedefaultseppunct}\relax
\EndOfBibitem
\bibitem[Carter and Hartzell(2020)]{carter_effect_2020}
D.~Carter and C.~Hartzell, \emph{Journal of Electrostatics}, 2020, \textbf{107}, 103475\relax
\mciteBstWouldAddEndPuncttrue
\mciteSetBstMidEndSepPunct{\mcitedefaultmidpunct}
{\mcitedefaultendpunct}{\mcitedefaultseppunct}\relax
\EndOfBibitem
\bibitem[Waitukaitis and Jaeger(2013)]{waitukaitis_situ_2013}
S.~R. Waitukaitis and H.~M. Jaeger, \emph{Review of Scientific Instruments}, 2013, \textbf{84}, 025104\relax
\mciteBstWouldAddEndPuncttrue
\mciteSetBstMidEndSepPunct{\mcitedefaultmidpunct}
{\mcitedefaultendpunct}{\mcitedefaultseppunct}\relax
\EndOfBibitem
\bibitem[Forward \emph{et~al.}(2009)Forward, Lacks, and Mohan~Sankaran]{forward_methodology_2009}
K.~M. Forward, D.~J. Lacks and R.~Mohan~Sankaran, \emph{Journal of Electrostatics}, 2009, \textbf{67}, 178--183\relax
\mciteBstWouldAddEndPuncttrue
\mciteSetBstMidEndSepPunct{\mcitedefaultmidpunct}
{\mcitedefaultendpunct}{\mcitedefaultseppunct}\relax
\EndOfBibitem
\bibitem[Keithley(1998)]{keithley_low_1998}
J.~F. Keithley, \emph{Low {Level} {Measurements}. {Precision} {DC} {Current}, {Voltage} and {Resistance} {Measurements}}, Keithley Instruments, 1998\relax
\mciteBstWouldAddEndPuncttrue
\mciteSetBstMidEndSepPunct{\mcitedefaultmidpunct}
{\mcitedefaultendpunct}{\mcitedefaultseppunct}\relax
\EndOfBibitem
\bibitem[Whitby(1978)]{whitby_physical_1978}
K.~T. Whitby, \emph{Sulfur in the {Atmosphere}}, Elsevier, 1978, pp. 135--159\relax
\mciteBstWouldAddEndPuncttrue
\mciteSetBstMidEndSepPunct{\mcitedefaultmidpunct}
{\mcitedefaultendpunct}{\mcitedefaultseppunct}\relax
\EndOfBibitem
\bibitem[Chen and Fryrear(2001)]{chen_aerodynamic_2001}
W.~Chen and D.~W. Fryrear, \emph{Journal of Sedimentary Research}, 2001, \textbf{71}, 365--371\relax
\mciteBstWouldAddEndPuncttrue
\mciteSetBstMidEndSepPunct{\mcitedefaultmidpunct}
{\mcitedefaultendpunct}{\mcitedefaultseppunct}\relax
\EndOfBibitem
\bibitem[Watson \emph{et~al.}(2010)Watson, Chow, Chen, and Wang]{watson_measurement_2010}
J.~Watson, J.~Chow, L.-W.~A. Chen and X.~Wang, \emph{Measurement system evaluation for fugitive dust emissions detection and quantification}, Desert Research Institute, 2010\relax
\mciteBstWouldAddEndPuncttrue
\mciteSetBstMidEndSepPunct{\mcitedefaultmidpunct}
{\mcitedefaultendpunct}{\mcitedefaultseppunct}\relax
\EndOfBibitem
\bibitem[Perry \emph{et~al.}(2008)Perry, Green, Green, and Perry]{perry_perrys_2008}
R.~H. Perry, D.~W. Green, D.~W. Green and R.~H. Perry, \emph{Perry's chemical engineers' handbook.}, McGraw-Hill Professional, New York, 8th edn, 2008\relax
\mciteBstWouldAddEndPuncttrue
\mciteSetBstMidEndSepPunct{\mcitedefaultmidpunct}
{\mcitedefaultendpunct}{\mcitedefaultseppunct}\relax
\EndOfBibitem
\bibitem[Balevičius and Miškinis(2020)]{balevicius_air_2020}
R.~Balevičius and P.~Miškinis, \emph{Mokslas – Lietuvos ateitis / Science – Future of Lithuania}, 2020, \textbf{12}, 1--4\relax
\mciteBstWouldAddEndPuncttrue
\mciteSetBstMidEndSepPunct{\mcitedefaultmidpunct}
{\mcitedefaultendpunct}{\mcitedefaultseppunct}\relax
\EndOfBibitem
\bibitem[Venczel \emph{et~al.}(2021)Venczel, Bognár, and Veress]{venczel_temperature-dependent_2021}
M.~Venczel, G.~Bognár and A.~Veress, \emph{Processes}, 2021, \textbf{9}, 331\relax
\mciteBstWouldAddEndPuncttrue
\mciteSetBstMidEndSepPunct{\mcitedefaultmidpunct}
{\mcitedefaultendpunct}{\mcitedefaultseppunct}\relax
\EndOfBibitem
\bibitem[Ruzer and Harley(2012)]{ruzer_aerosols_2012}
L.~S. Ruzer and N.~H. Harley, \emph{Aerosols {Handbook}: {Measurement}, {Dosimetry}, and {Health} {Effects}, {Second} {Edition}}, CRC Press, 2012\relax
\mciteBstWouldAddEndPuncttrue
\mciteSetBstMidEndSepPunct{\mcitedefaultmidpunct}
{\mcitedefaultendpunct}{\mcitedefaultseppunct}\relax
\EndOfBibitem
\bibitem[Kok and Lacks(2009)]{kok_electrification_2009}
J.~F. Kok and D.~J. Lacks, \emph{Physical Review E}, 2009, \textbf{79}, 051304\relax
\mciteBstWouldAddEndPuncttrue
\mciteSetBstMidEndSepPunct{\mcitedefaultmidpunct}
{\mcitedefaultendpunct}{\mcitedefaultseppunct}\relax
\EndOfBibitem
\bibitem[Smallenburg(2022)]{smallenburg_efficient_2022}
F.~Smallenburg, \emph{Efficient event-driven simulations of hard spheres}, 2022\relax
\mciteBstWouldAddEndPuncttrue
\mciteSetBstMidEndSepPunct{\mcitedefaultmidpunct}
{\mcitedefaultendpunct}{\mcitedefaultseppunct}\relax
\EndOfBibitem
\bibitem[Lacks and Levandovsky(2007)]{lacks_effect_2007}
D.~J. Lacks and A.~Levandovsky, \emph{Journal of Electrostatics}, 2007, \textbf{65}, 107--112\relax
\mciteBstWouldAddEndPuncttrue
\mciteSetBstMidEndSepPunct{\mcitedefaultmidpunct}
{\mcitedefaultendpunct}{\mcitedefaultseppunct}\relax
\EndOfBibitem
\bibitem[Arason \emph{et~al.}(2011)Arason, Bennett, and Burgin]{arason_charge_2011}
P.~Arason, A.~J. Bennett and L.~E. Burgin, \emph{Journal of Geophysical Research: Solid Earth}, 2011, \textbf{116}, 1--15\relax
\mciteBstWouldAddEndPuncttrue
\mciteSetBstMidEndSepPunct{\mcitedefaultmidpunct}
{\mcitedefaultendpunct}{\mcitedefaultseppunct}\relax
\EndOfBibitem
\bibitem[McNutt and Williams(2010)]{mcnutt_volcanic_2010}
S.~R. McNutt and E.~R. Williams, \emph{Bulletin of Volcanology}, 2010, \textbf{72}, 1153--1167\relax
\mciteBstWouldAddEndPuncttrue
\mciteSetBstMidEndSepPunct{\mcitedefaultmidpunct}
{\mcitedefaultendpunct}{\mcitedefaultseppunct}\relax
\EndOfBibitem
\bibitem[O'Hara \emph{et~al.}(2024)O'Hara, Marsden, Reid, and Aplin]{ohara_faraday_2024}
T.~F. O'Hara, G.~Marsden, D.~P. Reid and K.~L. Aplin, \emph{Faraday {Cup} {Measurements} of {Triboelectrically} {Charged} {Granular} {Material}: {A} {Modular} {Interpretation} {Methodology}}, 2024, \url{https://www.doi.org/10.18126/n3dx-dh16}\relax
\mciteBstWouldAddEndPuncttrue
\mciteSetBstMidEndSepPunct{\mcitedefaultmidpunct}
{\mcitedefaultendpunct}{\mcitedefaultseppunct}\relax
\EndOfBibitem
\bibitem[O'Hara(2024)]{ohara_vb22224drop_modelling_and_fitting_2024}
T.~O'Hara, \emph{vb22224/{Drop}\_Modelling\_and\_Fitting: {Faraday} {Cup} {Modelling} and {Fitting}}, 2024, \url{https://zenodo.org/doi/10.5281/zenodo.13774132}\relax
\mciteBstWouldAddEndPuncttrue
\mciteSetBstMidEndSepPunct{\mcitedefaultmidpunct}
{\mcitedefaultendpunct}{\mcitedefaultseppunct}\relax
\EndOfBibitem
\bibitem[O'Hara and Reid(2024)]{ohara_vb22224simplified_edmd_triboelectric_2024}
T.~O'Hara and D.~Reid, \emph{vb22224/{Simplified}\_EDMD\_Triboelectric \_Model: {Simplified} {EDMD} {Triboelectric} {Models}}, 2024, \url{https://zenodo.org/doi/10.5281/zenodo.13774615}\relax
\mciteBstWouldAddEndPuncttrue
\mciteSetBstMidEndSepPunct{\mcitedefaultmidpunct}
{\mcitedefaultendpunct}{\mcitedefaultseppunct}\relax
\EndOfBibitem
\end{mcitethebibliography}
\bibliographystyle{rsc} 

\end{document}